\documentclass[aps,prl,amsmath,preprintnumbers,twocolumn]{revtex4}

\usepackage{epsfig}
\usepackage{graphics}
\usepackage{latexsym}
\usepackage{amsmath}
\usepackage{amssymb}
\usepackage{rotating}
\usepackage{subfigure}
\usepackage{bm}
\usepackage{color}
\usepackage{amsfonts}
\usepackage{amsmath}
\usepackage{mathtools}

\usepackage{enumitem}

\begin{document}

\title{Chiral extrapolation of nucleon axial charge $g_A$ in effective field theory}
\author{Hongna Li$^{1,2}$}
\author{P. Wang$^{1,3}$}

\affiliation{$^1$Institute of High Energy Physics, CAS, Beijing 100049, China}
\affiliation{$^2$College of Physics, Jilin University, Changchun, Jilin 130012, China}
\affiliation{$^3$Theoretical Physics Center for Science Facilities, CAS, Beijing 100049,
China}
\begin{abstract}
The extrapolation of nucleon axial charge $g_A$ is investigated within the framework of heavy baryon chiral effective field theory. The intermediate octet and decuplet baryons are included in the one loop calculation. Finite range regularization is applied to improve the convergence in the quark-mass expansion. The lattice data from three different groups are used for the extrapolation. At physical pion mass, the extrapolated $g_A$ are all smaller than the experimental value.

\end{abstract}
\maketitle


The nucleon axial charge, $g_A$, is a fundamental property of the nucleon, which reveals how the up and down quark intrinsic spin contribute to the spin of the proton and neutron, governing $\beta$ decay and providing a quantitative measure of spontaneous chiral symmetry
breaking in low energy hadronic physics.
The axial charge $g_A$ is of great importance to any further calculation of hadron structure.

The axial charge $g_A$ is defined as the
axial vector form factor at zero four-momentum transfer,
$g_A=G_A(0)$. The axial vector form factor is
given by the nucleon matrix element of the axial vector
current, $A_\mu ^a = \overline \psi  {\gamma _\mu }{\gamma _5}\left( {{{{\tau ^a}} \mathord{\left/ {\vphantom {{{\tau ^a}} 2}} \right.
\kern-\nulldelimiterspace} 2}} \right)\psi $,
with $u$, $d$ quark doublet $\psi$,
$\left\langle {N(p',s')} \right|A_\mu ^3\left| {N(p,s)} \right\rangle  = i\bar u(p',s')[{\gamma _\mu }{\gamma _5}{G_A}({q^2}) + \frac{{{q_\mu }}}{{2{M_N}}}{\gamma _5}{G_P}({q^2})]\frac{{{\tau ^3}}}{2}u(p,s)$,
where $G_P$ is the induced pseudoscalar form factor, $\tau^a$ is
an isospin Pauli matrix, and $q_\mu=p_\mu'-p_\mu$ is the momentum transfer.
At zero momentum transfer, the axial charge $g_A$ is the spin difference between $u$ and $d$ quark
in proton, i.e.
\begin{equation}
g_A=\Delta u - \Delta d.
\end{equation}

Experimentally, $g_A$ has been obtained very
precisely through neutron $\beta$ decay, with the Particle Data Group value $g_A = 1.27\pm 0.003$~\cite{Beringer:1900zz}.
Theoretically, there are many calculations in different methods, such as the cloudy bag model \cite{Bass:2009ed}, the perturbative chiral quark model \cite{Liu:2014owa}, the relativistic constituent quark model \cite{Boffi:2001zb}, Schwinger-Dyson formalism \cite{Yamanaka:2014lva}, chiral perturbation theory \cite{Bernard:2007zu}, etc. There are also many lattice simulations of axial
charge \cite{Bratt:2010jn,Alexandrou:2010hf,Capitani:2012gj,Edwards:2005ym,Ohta:2013qda,Yamazaki:2008py}. Due to the limitation of the computing ability, all the simulations of $g_A$ are at
large quark mass. The obtained $g_A$ at large quark mass are smaller than the experimental data.
Therefore, it is interesting to see how the axial charge $g_A$ changes at low pion mass.

In this paper, we will extrapolate nucleon axial charge $g_A$ in the framework of heavy baryon chiral perturbation theory with finite range regularization (FRR). FRR has been applied in the extrapolation
of nucleon mass, magnetic form factors, strange form factors, charge radii, first moments, etc \cite{Young:2002ib,Leinweber:2003dg,Wang:2007iw,Wang:2010hp,Allton:2005fb,Armour:2008ke,Hall:2013oga,Leinweber:2004tc,Wang:1900ta,Wang:2012hj,Wang:2013cfp,Hall:2013dva,Wang:2015sdp,Li:2015exr,Wang:2008vb}. It is proved that FRR can provide a good convergent behaviour of pion mass expansion. Therefore, it is expected to have a good  description of the pion mass dependence of axial charge $g_A$ in wide range of pion mass.

The lowest-order chiral
Lagrangian including the octet and decuplet baryons is expressed as
\begin{eqnarray}
{L_{v}}&=&iTr\overline{{B}}{_{v}}\left( {v\cdot {\cal D}}\right) {B_{v}}+2DTr
\overline{{B}}{_{v}}S_{v}^{\mu }\left\{ {{A_{\mu }},{B_{v}}}\right\}
\nonumber \\
&&+2FTr\overline{{B}}{_{v}}S_{v}^{\mu }\left[ {{A_{\mu }},{B_{v}}}\right]
-i\overline{T}_{v}^{\mu }\left( {v\cdot {\cal D}}\right) {T_{v\mu }}
\nonumber \\
&&+{\cal C}\left( {\overline{T}_{v}^{\mu }{A_{\mu }}{B_{v}}+\overline{{B}}_{v}{A_{\mu
}}T_{v}^{\mu }}\right) \, ,
\end{eqnarray}
where $S_{v}^{\mu }$ is the covariant spin operator defined as
\begin{equation}
S_{v}^{\mu }=\frac{i}{2}{\gamma ^{5}}{\sigma ^{\mu \nu }}{v_{\nu }}\, .
\end{equation}%
Here, $v^{\nu }$ is the nucleon four velocity.  In the rest frame, we
have $ v^{\nu }=(1,0,0,0)$). $D$, $F$ and ${\cal C}$ are the standard
$SU(3)$-flavour coupling constants.
\begin{figure}[t]
\includegraphics[width=7cm]{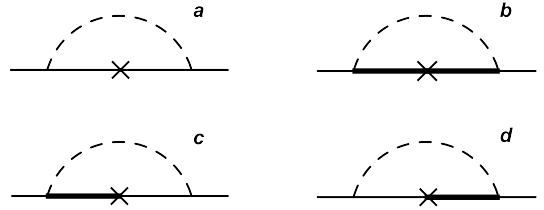}
\caption{The one-loop Feynman diagrams for calculating the quark
contribution to the proton spin. The thin and thick solid lines are
for the octet and decuplet baryons, respectively. }
\label{fig:1}
\end{figure}

According to the Lagrangian, the one-loop Feynman diagrams, which
contribute to axial charge $g_A$ of the proton, are plotted in
Fig.~1.  The axial charge is the spin difference between $u$ and $d$ quark.
The contribution of  $u$-, $d$-quark sector to the proton spin,
from Fig.~\ref{fig:1}a, are expressed as
\begin{eqnarray}
\Delta u^{a}&=& \left [ C_{N\pi }\, I_{2\pi }^{NN}+C_{\Sigma K}\, I_{2K}^{N\Sigma
}+C_{\Lambda \Sigma K}\, I_{5K}^{N\Lambda \Sigma } \right . \nonumber \\
&&\left . + C_{N\eta }\, I_{2\eta }^{NN}\right ]s_{u} \, , \\
\Delta d^{a}&=& \left [ \frac{7}{2}C_{N\pi }\, I_{2\pi }^{NN} +
  \frac{1}{5} C_{\Sigma K}\, I_{2K}^{N\Sigma } - C_{\Lambda \Sigma
    K}\, I_{5K}^{N\Lambda \Sigma }
  \right . \nonumber \\
&&\left . -\frac{1}{4} C_{N\eta }\, I_{2\eta }^{NN}\right ] s_{d} ,
\end{eqnarray}
where the coefficients, $C$ are expressed as
\begin{eqnarray}
C_{N\pi}
&=&-\frac{(D+F)^2}{288\,\pi^3\, f_\pi^2} \, , \\
C_{\Sigma K}
&=&-\frac{5(D-F)^{2}}{288\,\pi^3\, f_\pi^2} \, , \\
C_{\Lambda \Sigma K}
&=&\frac{(D-F)\,(D+3F)}{288\,\pi^3\, f_\pi^2} \, ,\\
C_{N\eta }
&=&-\frac{2}{3}\, \frac{(3F-D)^{2}}{288\,\pi^3\, f_\pi^2} \, .
\end{eqnarray}

The tree level contributions to the proton spin from $u$ and $d$ quark of intermediate octet baryons
are used in the above formulas.
For example, for the intermediate proton and neutron, their spins are
expressed as
\begin{equation}
s_p = \frac{4}{3} s_u - \frac{1}{3} s_d\, , ~~~ s_n = \frac{4}{3} s_d -
\frac{1}{3} s_u \, .
\end{equation}
$s_u$ and $s_d$ are the single quark spin of $u$ and $d$ quark. With the $SU(2)$ symmetry,
$s_u = s_d = s_q$ and $s_q$ can be written as
\begin{equation}\label{sq}
s_{q}\left( m_{\pi }^{2}\right)
=c_{0}+c_{2}m_{\pi }^{2}+c_{4}m_{\pi }^{4},
\end{equation}
where $c_0$, $c_2$ and $c_4$ are the low energy constants.

The contribution of  $u$-, $d$-quark sector to the proton spin,
described by diagram (b) of Fig.~1, are expressed as
\begin{eqnarray}
\Delta u^{b}&=& \left [ C_{\Delta\pi}\, I^{N\Delta}_{2\pi} + C_{\Sigma^*
    K}\, I^{N\Sigma^*}_{2K} \right ]\, s_u \, , \\
\Delta d^{b}&=& \left [ \frac{2}{7}\, C_{\Delta\pi}\, I^{N\Delta}_{2\pi}
  + \frac{1}{5}\, C_{\Sigma^* K}\, I^{N\Sigma^*}_{2K}\right ]\, s_d.
\end{eqnarray}
where the coefficients $C_{\Delta\pi}$ and $C_{\Sigma^* K}$ are
\begin{eqnarray}
C_{\Delta\pi}  &=& \frac{35\, {\cal C}^2}{648\, \pi^3\, f_\pi^2} \, , \\
C_{\Sigma^* K} &=& \frac{5}{28}\, C_{\Delta\pi} \, .
\end{eqnarray}
Similar as in the case of octet intermediate state, the tree level quark contributions to the
spin of decuplet baryons are also used.  For example
\begin{equation}
s_{\Delta^+} = 2\, s_u + s_d\, , ~~~ s_{\Sigma^{*-}} = 2\, s_d + s_s \, .
\end{equation}

Diagrams (c) and (d) of Fig.~1 provide contributions from intermediate
states involving an octet-decuplet transition.  The $u$-, $d$-quark-sector
contribution to the proton spin from these diagrams are expressed as
\begin{eqnarray}
\Delta u^{c+d} &=& \left [ C_{N\Delta\pi}\, I^{N\Delta}_{3\pi}
+ C_{\Sigma\Sigma^* K}\, I^{N\Sigma\Sigma^*}_{5K}
+ C_{\Lambda\Sigma^* K}\, I^{N\Lambda\Sigma^*}_{5K} \right ] \nonumber \\
&& \times  s_u \, , \\
\Delta d^{c+d}&=& \left [ -C_{N\Delta \pi }\, I_{3\pi }^{N\Delta }
+\frac{1}{5}\, C_{\Sigma \Sigma ^{\ast }K}\, I_{5K}^{N\Sigma \Sigma
  ^{\ast }} \right . \nonumber \\
&&\qquad \left . -C_{\Lambda \Sigma ^{\ast }K}\, I_{5K}^{N\Lambda \Sigma ^{\ast
}}\right ]\, s_{d}\, .
\end{eqnarray}
where
\begin{eqnarray}
C_{N\Delta\pi}        &=& -\frac{(D+F)\,{\cal C}}{27\, \pi^3\, f_\pi^2} \, , \\
C_{\Sigma\Sigma^* K}  &=& -\frac{5}{8}\, \frac{(D-F) \,{\cal C}}{27\, \pi^3\, f_\pi^2} \, , \\
C_{\Lambda\Sigma^* K} &=& -\frac{1}{8}\, \frac{(D+3F)\,{\cal C}}{27\, \pi^3\, f_\pi^2} \, .
\end{eqnarray}
The integrals in the above equations, $I^{\alpha\beta}_{2j}$,
$I^{\alpha\beta\gamma}_{5j}$ and $I^{\alpha\beta}_{3j}$ are defined in
Ref.~\cite{Wang:2007iw}.

The total $u$-, $d$- quark sector contributions to the spin of the  proton are written as
\begin{eqnarray}
\Delta {u}&=&Z[\frac{4}{3} ({c_{0}+c_{2}m_{\pi }^{2}+c_{4}m_{\pi }^{4}})
          +\Delta u^{a}+\Delta u^{b}+\Delta u^{c+d}] , \nonumber\\
\Delta {d}&=&Z[-\frac{1}{3} (c_{0}+c_{2}m_{\pi }^{2}+c_{4}m_{\pi }^{4})
+\Delta d^{a}+\Delta d^{b}+\Delta d^{c+d}],\nonumber\\
&&
\end{eqnarray}
where $Z$ is the wave function renormalization constant, expressed as
\begin{eqnarray}
{Z} &=& 1/[1 + \frac{1}{{48{\pi ^3}f_\pi ^2}}(\beta _\pi ^{NN}I_{2\pi }^{NN} + \beta _\pi ^{N\Delta }I_{2\pi }^{N\Delta } + \beta _{\rm{K}}^{N\Lambda }I_{2{\rm{K}}}^{N\Lambda } \nonumber \\
&&+ \beta _{\rm{K}}^{N\Sigma }I_{2{\rm{K}}}^{N\Sigma } + \beta _{\rm{K}}^{N{\Sigma ^ * }}I_{2{\rm{K}}}^{N{\Sigma ^ * }} + \beta _\eta ^{NN}I_{2\eta }^{NN})]
\end{eqnarray}
The above coefficients are expressed as
\begin{eqnarray}
\beta _\pi ^{NN} &=& \frac{9}{4}{\left( {D + F} \right)^2},\quad \beta _\pi ^{N\Delta } = 2{{\cal C}^2} \nonumber\\
\beta _{\rm{K}}^{N\Lambda } &=& \frac{1}{4}{\left( {3F + D} \right)^2},\quad
\beta _{\rm{K}}^{N\Sigma } = \frac{9}{4}{\left( {D - F} \right)^2} \nonumber\\
\beta _{\rm{K}}^{N{\Sigma ^ * }} &=& \frac{1}{2}{{\cal C}^2},\qquad \beta _\eta ^{NN} = \frac{1}{4}{\left( {3F - D} \right)^2}
\end{eqnarray}

The $K-$ and $\eta-$ meson masses have relationships with the
pion mass as
\begin{eqnarray}
m_K^2 &=& \frac{1}{2}m_\pi ^2 + {\left. {m_K^2} \right|_{phy}} - \frac{1}{2}{\left. {m_\pi ^2} \right|_{phy}},\\
m_\eta ^2 &=& \frac{1}{3}m_\pi ^2 + {\left. {m_\eta ^2} \right|_{phy}} - \frac{1}{3}{\left. {m_\pi ^2} \right|_{phy}}.
\end{eqnarray}
This enable a direct relationship between the nucleon axial charge and the pion mass.
By fitting the lattice data with different pion mass, we can get the low energy constants
$c_0$, $c_2$ and $c_4$.

In our calculation, the one-gluon-exchange is also included.
Although it lies outside the framework of chiral effective field theory,
the effect of one-gluon-exchange (OGE) is particularly important for
spin dependent quantities. Hogaason and Myhrer~\cite{Hogaasen:1987nj}
showed that the incorporation of the exchange current correction arising from the effective
one-gluon-exchange (OGE) force shifts the tree-level non-singlet charge, $g_A$, from
$\frac{5}{3} s_q$ to $\frac{5}{3} s_q - G$, where $G$ is about 0.05. The OGE correction shifts
the tree-level singlet charge $g_0$ from  $s_q$ to $s_q - 3G$.
In other words, the spin of each constituent quark gain a OGE correction $-G$\ at tree level.

In the numerical calculations, the couplings constant $D$ and $F$ are
$D=0.8$, $F=0.46$. The decuplet coupling ${\cal C}$ is chosen to be
$-1.2$ as in Ref.~\cite{Jenkins:1992pi}.  The regulator in the integrals is chosen
to be of a dipole form
\begin{equation}
u(k) =\frac{1}{\left(1+k^2/\Lambda^2\right)^2} \, ,
\end{equation}
with $\Lambda=0.8$ GeV. This regulator has been used in our previous study
of nucleon mass, magnetic form factors, strange form factors, charge radii, first moments, etc
\cite{Young:2002ib,Leinweber:2003dg,Wang:2007iw,Wang:2010hp,Allton:2005fb,Armour:2008ke,Hall:2013oga,Leinweber:2004tc,Wang:1900ta,Wang:2012hj,Wang:2013cfp,Hall:2013dva,Wang:2015sdp,Li:2015exr,Wang:2008vb}.

\begin{figure}[t]
\includegraphics[width=8.5cm]{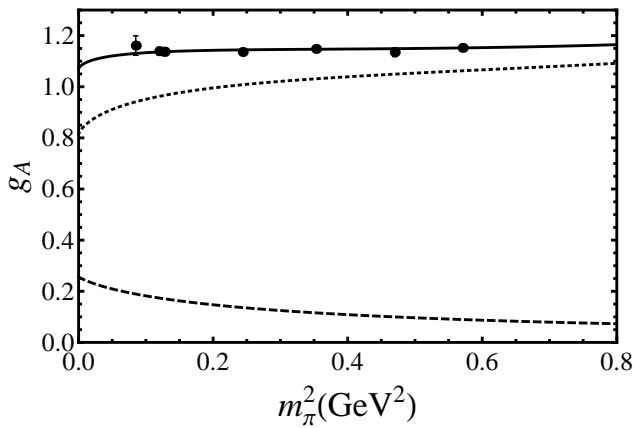}
\caption{$g_A$ fitted by the lattice data of Ref.~\cite{Bratt:2010jn} at $\Lambda=0.8$. The dotted, dashed and solid lines are for the tree level, loop and total contribution, respectively. }
\label{fig:2}
\end{figure}

In Fig.~\ref{fig:2}, the pion mass dependence of $g_A$ with $\Lambda = 0.8$ GeV is shown for lattice data of Ref.~\cite{Bratt:2010jn}. The dotted, dashed and solid lines are for tree level, loop and total contribution, respectively.
At large pion mass, the axial charge $g_A$ changes little. At small pion mass, $g_A$ decreases with
the decreasing pion mass. Compared with the pion mass dependence of proton magnetic form factors\cite{Wang:2007iw,Wang:2012hj}, at low pion mass, the curvature is small and opposite. This is because the leading diagram in the case of magnetic form factor has no contribution for $g_A$.
At physical pion mass, the extrapolated $g_A$ is 1.10, which is smaller than the experimental value 1.27.

\begin{figure}[t]
\includegraphics[width=8.5cm]{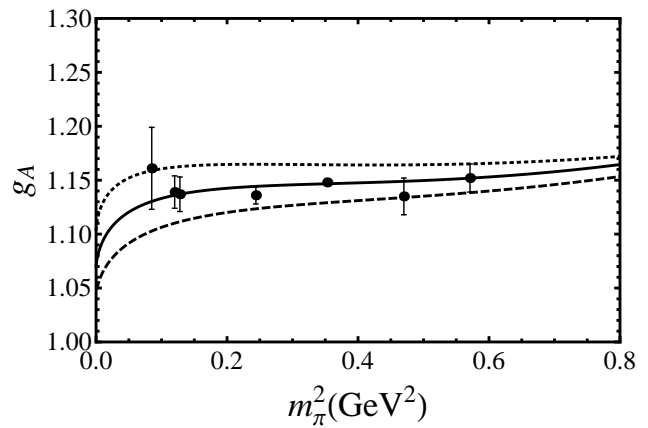}
\caption{Error band of $g_A$ fitted by the lattice data of Ref.~\cite{Bratt:2010jn}. The upper (dotted) line is for the upper limit with $g_A=\Delta u$($\Lambda=0.6$ GeV)$-\Delta d$($\Lambda=1.0$ GeV). The middle (solid) line is for the central value of $g_A$ ($\Lambda=0.8$ GeV). The lower (dashed) line is for the lower limit with $g_A=\Delta u$($\Lambda=1.0$ GeV)$-\Delta d$($\Lambda=0.6$ GeV).}
\label{fig:3}
\end{figure}

To provide an estimate of the uncertainty in these results, we vary
the regulator parameter, $\Lambda$, governing the size of meson
cloud contributions to proton structure. Considering $\Lambda = 0.8
\pm 0.2 $ GeV, the obtained low energy constants $c_0$, $c_2$, $c_4$
as well as the quark spin at physical pion mass are listed in Table~\ref{tab1}.
By varying $\Lambda$, we can provide an error bar for $g_A$. For example,
the highest and lowest $g_A$ at physical pion mass are 1.14 ($0.805 - (-0.333)$)
and 1.07 ($0.772 - (-0.302)$). From the table, one can see that the loop/tree contribution
increases/decreases with the increasing $\Lambda$.
The highest and lowest value of $g_A$ versus pion mass as well as the central value
of $g_A$ are shown in Fig.~\ref{fig:3}. It is clear that the extrapolated $g_A$ with error bar
is still smaller than experimental value.

\begin{table*}[tbp]
\caption{The parameters fitted by the lattice data of Ref.~\cite{Bratt:2010jn} and the obtained quark spin of the proton at physical pion mass for the regulator parameter $\Lambda=0.6,~0.8,~1.0$ GeV.}
\label{tab1}
\begin{ruledtabular}
\begin{tabular}{c|ccc|ccccccc}
$\Lambda$ (GeV)& $c_0$ & $c_2$ (GeV$^{-2}$)&  $c_4$ (GeV$^{-4}$)& $Z$   & $\Delta u$ & $\Delta d$ & $g_A$ & tree & loops  \\ \hline
0.6 & 0.74 & -0.04 & 0.04 & 0.84 & 0.80 & -0.30 & 1.107 & 0.99 & 0.12  \\
0.8 & 0.77 & -0.09 & 0.07 & 0.71 & 0.79 & -0.32 & 1.104 & 0.87 & 0.23  \\
1.0 & 0.81 & -0.12 & 0.09 & 0.58 & 0.77 & -0.33 & 1.106 & 0.75 & 0.36
\end{tabular}
\end{ruledtabular}
\end{table*}

\begin{figure}[t]
\includegraphics[width=8.5cm]{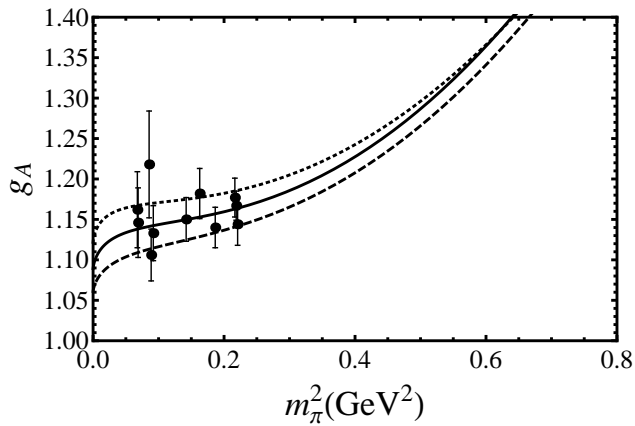}
\caption{Error band of $g_A$ fitted by the lattice data of Ref.~\cite{Alexandrou:2010hf}. The upper (dotted) line is for the upper limit with $g_A=\Delta u$($\Lambda=0.6$ GeV)$-\Delta d$($\Lambda=1.0$ GeV). The middle (solid) line is for the central value of $g_A$ ($\Lambda=0.8$ GeV). The lower (dashed) line is for the lower limit with $g_A=\Delta u$($\Lambda=1.0$ GeV)$-\Delta d$($\Lambda=0.6$ GeV).}
\label{fig:4}
\end{figure}

\begin{figure}[t]
\includegraphics[width=8.5cm]{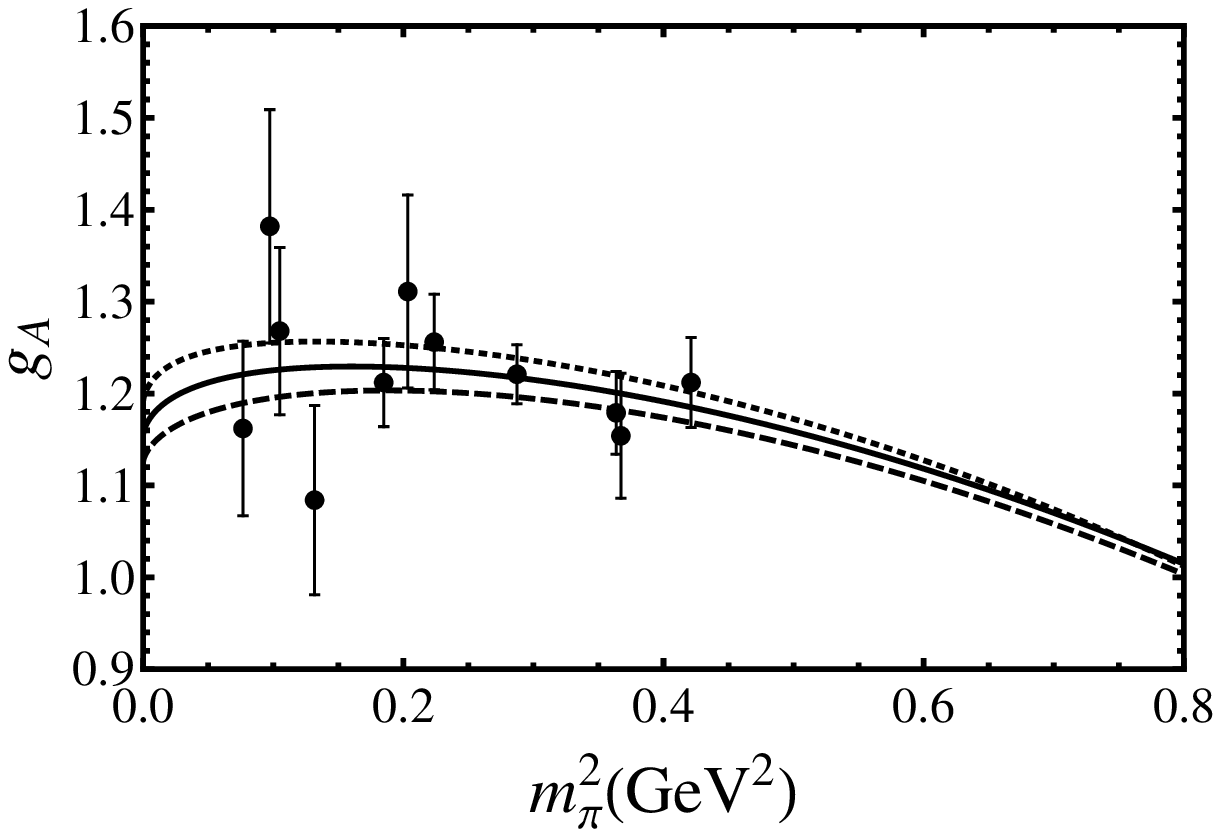}
\caption{Error band of $g_A$ fitted by the lattice data of Ref.~\cite{Capitani:2012gj}. The upper (dotted) line is for the upper limit with $g_A=\Delta u$($\Lambda=0.6$ GeV)$-\Delta d$($\Lambda=1.0$ GeV). The middle (solid) line is for the central value of $g_A$ ($\Lambda=0.8$ GeV). The lower (dashed) line is for the lower limit with $g_A=\Delta u$($\Lambda=1.0$ GeV)$-\Delta d$($\Lambda=0.6$ GeV).}
\label{fig:5}
\end{figure}

There are also other lattice groups simulating the axial charge $g_A$.
Fig.~\ref{fig:4} and Fig.~\ref{fig:5} are results for the lattice data from Refs.~\cite{Alexandrou:2010hf} and \cite{Capitani:2012gj}. Same as in Fig.~\ref{fig:3}, the lines in the middle are for $\Lambda = 0.8$ GeV. The upper and lower lines are obtained by varying $\Lambda$ from 0.6 to 1 GeV. The extrapolated $g_A$ from Ref.~\cite{Alexandrou:2010hf} at physical pion mass is $1.12^{+0.03}_{-0.03}$. The lattice data from
Ref.~\cite{Capitani:2012gj} varied a lot with the change of the pion mass though the extrapolated $g_A$ at physical pion mass is a little larger than the other two lattice groups.
At large pion mass, Fig.~\ref{fig:4} and Fig.~\ref{fig:5} show that $g_A$ changes quickly with the increasing pion mass for the data of ETMC and
data from Ref.~\cite{Capitani:2012gj}. This is because different from the data of LHPC, there is no constraint from these lattice date at large pion mass.
Overall, one can see the results from different lattice groups are comparable and all
the extrapolated $g_A$ at physical pion mass are smaller than the experimental value with the
error bar. The obtained results with central $\Lambda = 0.8$ GeV for these three lattice groups are
listed in Table~\ref{tab2}.

\begin{table*}[tbp]
\caption{The parameters fitted by three group lattice data~\cite{Bratt:2010jn,Alexandrou:2010hf,Capitani:2012gj} and the obtained quark spin of the proton at physical pion mass for the regulator parameter $\Lambda=0.8$ GeV.}
\label{tab2}
\begin{ruledtabular}
\begin{tabular}{c|ccc|ccccccc}
lattice data  & $c_0$ & $c_2$ (GeV$^{-2}$)&  $c_4$ (GeV$^{-4}$)& $Z$   & $\Delta u$ & $\Delta d$ & tree & loops & $g_A$ with error bar \\ \hline
Ref.~\cite{Bratt:2010jn}      & 0.77 & -0.09 & 0.07  & 0.71 & 0.79 & -0.32 & 0.87 & 0.23  & $1.10_{-0.03}^{+0.04}$\\
Ref.~\cite{Alexandrou:2010hf} & 0.78 & -0.21 & 0.60  & 0.71 & 0.80 & -0.32 & 0.88 & 0.24  & $1.12_{-0.03}^{+0.03}$\\
Ref.~\cite{Capitani:2012gj}   & 0.83 & -0.06 & -0.20 & 0.71 & 0.85 & -0.34 & 0.94 & 0.25  & $1.19_{-0.03}^{+0.04}$
\end{tabular}
\end{ruledtabular}
\end{table*}

In summary, we extrapolated the axial charge $g_A$ in chiral effective field theory
with finite range regularisation. The dipole regulator is used as our previous extrapolation
for nucleon mass, form factors, first moments, etc.
The lattice data are from three lattice groups where
the volume corrections are given explicitly. Different from the proton magnetic form factor,
the axial charge $g_A$ decreases with decreasing pion mass when $m_\pi$ is small.
The lattice data in wide pion mass range can be well described with the FRR chiral
effective field theory. At physical pion mass, the extrapolated
$g_A$ are comparable to each other and all of them are smaller than the experimental
value. To estimate the error bar for the extrapolation, we vary $\Lambda$ in the
regulator from 0.6 to 1 GeV. The up limit of the extrapolated $g_A$ at physical pion
mass is still smaller than the experimental value. We should mention that the volume
correction is given by the lattice groups. It will be interesting to extrapolate the lattice data
directly without volume correction in finite volume chiral effective field theory.

\section*{Acknowledgments}
This work is supported in part by the National Natural Science Foundation of China under 
Grant No. 11475186 and by Sino-German CRC 110 (NSFC Grant No.11621131001).

\end{document}